\documentclass
[aps,prb,superscriptaddress,showpacs,footinbib,twocolumn]{revtex4}%
\usepackage{amsmath}
\usepackage{graphicx}
\usepackage{amsfonts}
\usepackage{amssymb}
\usepackage{subfigure}
\usepackage{float}
\usepackage{color}%
\providecommand{\U}[1]{\protect\rule{.1in}{.1in}}
\providecommand{\U}[1]{\protect\rule{.1in}{.1in}}
\begin{document}
\title{Light-induced spin polarizations in quantum rings}
\author{Fateme K. Joibari}
\affiliation{Kavli Institute of NanoScience, Delft University of Technology, Delft, The Netherlands}
\author{Ya. M. Blanter}
\affiliation{Kavli Institute of NanoScience, Delft University of Technology, Delft, The Netherlands}
\author{Gerrit E. W. Bauer}
\affiliation{Kavli Institute of NanoScience, Delft University of Technology, Delft, The Netherlands}
\affiliation{Institute for Materials Research and WPI-AIMR, Tohoku University, Sendai, Japan}

\begin{abstract}
Non-resonant circularly polarized electromagnetic radiation can exert torques
on magnetization by the Inverse Faraday Effect (IFE). Here we discuss
the enhancement of IFE by spin-orbit interactions (SOI). We illustrate the
principle by studying a simple generic model system, {\em i.e.} the quasi-1D ring in
the presence of linear/cubic Rashba and Dresselhaus interactions. We combine
the classical IFE in electron plasmas that is known to cause persistent currents in the
plane perpendicular to the direction of the propagation of light with the
concept of current and spin-orbit-induced spin transfer torques. We calculate
light-induced spin polarization that in ferromagnets might give rise to
magnetization switching.
\end{abstract}
\maketitle


\section{Introduction}

The Faraday Effect (FE) describes the rotation of the plane of linearly
polarized light when it passes through a ferromagnet with magnetization
component parallel to the light vector. It is caused by the difference of phase
shifts of transmitted light in the two circular polarization states. The
Inverse Faraday Effect (IFE) is the ability of circularly polarized light to
exert torques on a magnetization, which can be interpreted in terms of an
effective light-induced magnetic field along its wave vector with the magnitude
proportional to the light intensity and the sign governed by its helicity. In
contrast to other photomagnetic effects neither FE nor IFE involve
absorption of photons, which makes them potentially very fast and therefore
interesting \textit{e.g}. for data storage technologies.

IFE was initially predicted by Pitaevskii\cite{IFE} and formulated in
terms of the dependence of the free energy on a time-dependent electric field.
After observation of IFE by van der Ziel \textit{et al.}\cite{IFE1},
Pershan \textit{et al}.\cite{Pershan} developed a microscopic theory
explaining IFE in terms of an optically-induced splitting of degenerate
spin levels, followed by thermal relaxation. They predicted that magnetization
$\vec{M}=V\lambda_{0}(2\pi c)^{-1}(I_{R}-I_{L}){\bf{e}}_{k}$ was created by the
circularly polarized light propagating in the ${\bf{e}}_{k}$ direction with
the intensity $I_{R(L)}$ of the right (left) handed circularly polarized light
component. Here $V$, $\lambda_{0},$ and $c$ are the Verdet constant,
the wavelength and the speed of the light, respectively.

Kimel \textit{et al.} demonstrated IFE in
$\text{DyFe}\text{O}_{3}$ by exciting magnetization dynamics with circularly
polarized laser pulses on fs time scales.\cite{kim} These and subsequent
experiments as reviewed in Ref. \onlinecite{rev-rasin} are not fully explained
by the theory presented by Pershan \textit{et al}.\cite{Pershan}, because thermal relaxation
does not occur at such short time scales. Subsequently, Stanciu \textit{et al.}%
\cite{stan} demonstrated that the perpendicular magnetization of GdFeCo thin
films can be switched on subpicosecond time scale. Vahaplar \textit{et
al.}\cite{vahaplar} modeled the switching process by multiscale calculations
of the magnetization dynamics \cite{kaz} with effective magnetic fields of the
order of $20\
\operatorname{T}%
$. However, the microscopic origin, magnitude and material dependence of these
fields remain unexplained.

The reciprocity between FE and IFE is not universally observed,\cite{Mikhaylovskiy} and was 
found by theory to break down in the presence of absorption.\cite{Battiato}  
Taguchi \textit{et al.} calculated the effect of terahertz electromagnetic
radiation on disordered metals with SOI.\cite{tatara} They found a light-induced magnetization, 
but at the cost of light absorption. This is in
contrast to the IFE phenomenology. Recently, strong effective magnetic fields
were calculated for magnetic semiconductors that are caused by the
spin-selective dynamical Stark effect.\cite{Qaium}

IFE has also been studied in classical plasmas, where it can be
explained in terms of the \O rsted magnetic fields generated by light-induced
circulating DC charge currents.\cite{kar,pla,pla1,pla2} Hertel investigated
this process for solid state electron plasmas.\cite{Hertel} He derived the
eddy currents and associated magnetic fields generated by time-dependent
circularly polarized light in a conducting metal film modeled as a
collisionless electron gas. Both currents and the related magnetic fields are
dissipationless and scale to second order in the electric field amplitude of
the circularly polarized light, in line with the microscopic theories for
IFE. However, the effects is too small to explain the
light-induced magnetization switching. Yoshino discussed dissipative
corrections to Hertel's theory.\cite{Yoshino}

Here, we pursue the concept that IFE is caused by light-induced DC
currents, but invoke the spin-orbit interaction (SOI) to explain the large
effective fields apparently at work. This perspective of IFE is motivated
by the linear current driven intrinsic spin torque in ferromagnets predicted
by Manchon and Zhang,\cite{man1,man2} who demonstrated that current in the
presence of SOI of the Rashba type produces an effective magnetic field
which is perpendicular to both an inversion symmetry-breaking electric field
and the current. The non-dissipative currents discussed above can be
interpreted as a reactive response to a light field, or as a ground state
property of the system in the presence of the light field, quite analogous to
persistent currents or diamagnetic response to magnetic field that can
be formulated as ground states in the presence of a vector
potential.\cite{Buttiker} The quantum-mechanical ground state nature of
the light-induced current in a $1D$ ring has been investigated by
Kibis.\cite{kibis} A possible route to a theory of IFE would be extending
the Kibis' approach to Hamiltonians with spin-orbit interactions. Rather than
focusing on the quantum mechanics of the generation of charge currents by the
light field, we concentrate here on the generation of effective magnetic
fields in the presence of circulating charge currents, while using the
Hertel's approach to estimate the magnitude of these currents for a
given light intensity. This is allowed in the high frequency limit in which
the length scale associated with the direct response is much smaller than the geometric
confinement or the spin-orbit precession length.

In order to establish the principle we focus here on a non-magnetic system
with spin-orbit interaction and in the presence of electron current bias
that is generated by circularly polarized light. For a magnetic sample, such polarization can exert
spin-orbit torques on magnetization. We focus on a simple yet
realistic model system in which the spin-orbit interaction Hamiltonian is well
known and analytical results can be achieved, \textit{viz}. a one-dimensional
(1D/single transverse mode) ring fabricated from a high-mobility
two-dimensional electron/hole gas (2DEG/2DHG) with Rashba and Dresselhaus SOI
interactions. SOI in a 2DEG that is linear in the wave vector is known to be
quite anomalous, causing \textit{e.g.} vanishing spin Hall effect by
impurity scattering.\cite{Inoue} Here we find that light-induced effective
fields in 1D rings with linear Rashba or Dresselhaus SOI also vanish, which
can be traced to the state independence of the equilibrium spin texture. The
holes of a 2DHG close to the valence band edge can also be described by Rashba
and Dresselhaus SOI interactions, but with a cubic dependence on the wave
vectors.\cite{Winkler,Zhang} A quantum ring containing a hole gas has an
out-of-plane state-dependent spin texture that indeed generates the
current-induced spin polarization. In a ferromagnet these would indeed induce
torques on a magnetization, thereby confirming our working hypothesis.

The remainder of the Article is organized as follows. We solve the problem
of a ground state in the presence of a given charge current by the method of
Lagrange multipliers as explained in Section \ref{meth}. In Section \ref{NoSOI}, 
we apply this method
to a simple case of rings in the absence of SOI, and discuss
the difference of the ground state current induced by Lagrangian multiplier
and the one induced by the magnetic field in a ring.
In Section
\ref{ring}, we discuss different models of SOI in more detail. In Sections
\ref{elRash} and \ref{elDress}, we address rings consisting of electrons in
the presence of linear Rashba or Dresselhaus SOI, respectively, in which the
current-induced spin polarization vanishes.
In Sections \ref{holesDress} and \ref{holesRash}, we continue with a p-doped quantum ring, in which a
current-induced polarization is generated by the cubic Dresselhaus or Rashba
SOI, respectively. We summarize our conclusions in Section \ref{con}.

\section{Method of Lagrange multipliers\label{meth}}

We are interested in the ground state of a conductor in the presence of
currents induced by an external perturbation such as electric field of light.
Rather than diagonalizing the Hamiltonian in the presence of electric
field,\cite{kibis} we calculate the ground state for given persistent
current. Hereby we lose some subtle non-perturbative quantum
effects\cite{kibis} beyond the current generation, which can be important when
microwave frequencies are tuned to lie between quantized states. This is
beyond the scope of this Article.

According to the current-density-functional theory\cite{Vignale} the ground state
energy of a system is a functional of the charge current distribution
$\mathbf{j}_{\mathrm{ext}}\left(  \mathbf{r}\right)  $. The minimum energy of
the system under the constraint of given $\mathbf{j}_{\mathrm{ext}}\left(
\mathbf{r}\right)  $ can be found by the method of Lagrange multipliers. Here
the Hamiltonian $H_{0}$ is augmented by the sum of the product of constraints
and Lagrange multipliers that in continuous systems becomes an integral. We
limit attention to non-interacting systems with single particle states
$\left\vert \Psi_{i}\right\rangle $ and occupation numbers $f_{i}\in\left\{
0,1\right\}  $ with $\sum_{i=1}^{\infty}f_{i}=N$ for a number of $N$
electrons. We may then express the constraint as
\begin{equation}
\label{co3D}\sum_{i}f_{i}\mathbf{j}_{i}\left(  \mathbf{R}\right)
=\mathbf{j}_{\mathrm{ext}}\left(  \mathbf{R}\right)  ,
\end{equation}
where the current operator $\mathbf{\hat{\jmath}}\left(  \mathbf{R}\right)  $
is defined in terms of the expectation value
\begin{align}
\mathbf{j}_{i}\left(  \mathbf{R}\right)   &  =\left\langle \Psi_{i}\right\vert
\mathbf{\hat{\jmath}}\left(  \mathbf{R}\right)  \left\vert \Psi_{i}%
\right\rangle \label{j1}\\
&  =\frac{e}{2}\int\Psi_{i}^{\ast}\left(  \mathbf{r}\right)  \left[
\mathbf{ {v}}\delta\left(  \mathbf{r}-\mathbf{R}\right)  +\delta\left(
\mathbf{r}-\mathbf{R}\right)  \mathbf{ {v}}\right]  \Psi_{i}\left(
\mathbf{r}\right)  d\mathbf{r}\label{j2}\\
&  =e\operatorname{Re}\Psi_{i}^{\ast}\left(  \mathbf{R}\right)  \mathbf{
{v}}\Psi_{i}\left(  \mathbf{R}\right)  \neq e\left\langle \Psi_{i}\right\vert
\mathbf{ {v}}\left\vert \Psi_{i}\right\rangle \label{j3}%
\end{align}
and $\mathbf{ {v}}$ is the velocity operator. The objective functional
under this constraint and the normalization condition $\left\langle \Psi
_{i}|\Psi_{i}\right\rangle =1$ is%
\begin{align}
F\left[  \left\{  \Psi_{i}\right\}  ,\mathbf{j}_{\mathrm{ext}}\right]   &
=\sum_{i}f_{i}\left(  \left\langle \Psi_{i}\right\vert H_{0}\left\vert
\Psi_{i}\right\rangle -\varepsilon_{i}\left(  \left\langle \Psi_{i}|\Psi
_{i}\right\rangle -1\right)  \right) \nonumber\\
&  +\int\mathbf{A}\left(  \mathbf{R}\right)  \cdot\left(  \mathbf{j}%
_{\mathrm{ext}}\left(  \mathbf{R}\right)  -\sum_{i}f_{i}\mathbf{j}_{i}\left(
\mathbf{R}\right)  \right)  d\mathbf{R}\ .
\end{align}
Here $\mathbf{A}$ is the Lagrange multiplier functional. Minimizing $F$,
{\em i.e.}\ $\delta F/\delta\Psi_{i}^{\ast}=0$, leads to the Schr\"{o}dinger
equation with the eigenfunctions $\left\vert \Psi_{i}\right\rangle $
corresponding to the Hamiltonian%
\begin{equation}
H=H_{0}-\int\mathbf{A}\left(  \mathbf{R}\right)  \cdot\mathbf{\hat{\jmath}%
}\left(  \mathbf{R}\right)  d\mathbf{R.} \label{h3d}%
\end{equation}
In the absence of spin-orbit interactions $\mathbf{j}_{i}=(e\hbar
/m)\operatorname{Im}\Psi_{i}^{\ast}\boldsymbol{\nabla}\Psi_{i}$ and%
\begin{equation}
H\left(  \mathbf{r},\mathbf{{p}}\right)  \rightarrow H_{0}\left(
\mathbf{r},\mathbf{{p}}-e\mathbf{A}\left(  \mathbf{r}\right)  \right)
-\frac{\hbar^{2}e^{2}\mathbf{A}^{2}\left(  \mathbf{r}\right)  }{2m} \ .
\label{HamG}%
\end{equation}
When the objective current density $\mathbf{j}_{\mathrm{ext}}\left(
\mathbf{r}\right)  $ is constant in space and time, the Lagrange function
$\mathbf{A}\left(  \mathbf{r}\right)  $ is a vector potential corresponding to
constant magnetic field, and the implementation of the charge current
constraint is equivalent to a gauge transformation. We note the close relation
with the current density functional theory,\cite{Vignale} in which effective
vector and scalar potentials are introduced to construct energy functionals of
charge and current densities. Finally, we observe that the time derivative of
the vector potential is the electric field, $\vec{E}=-d\mathbf{A}/dt$. 
Harmonic AC electric field therefore corresponds to a vector potential in the
same direction with the amplitude $A_{\omega}=-iE_{\omega}/\omega$ in
frequency space. The effect of finite $A_{\omega}$ in the DC limit
$\omega\rightarrow0$ is then equivalent to the transport response to
electric field that remains finite in a ballistic system. Alternatively, we
can associate the vector potential with applied magnetic field
inducing persistent ground state current, although it should be
kept in mind that when the current is generated by other means, our magnetic
field is a fictitious one.

\section{Single mode quantum rings without SOI \label{NoSOI}}

In the following we focus on quantum rings fabricated from 2DE(H)G in which
the charge carriers are confined normal to the plane by a potential $V(z)$ and
in the radial direction by an axially symmetric confining potential $U(r)$
centered at an effective radius $r=a$, but free to move along the azimuthal
direction along the unit vector $\mathbf{e}_{\varphi}$. In the envelope
function approximation with effective mass $m$ for electrons or (heavy)
holes:
\begin{equation}
H_{0}=\frac{{p}_{x}^{2}+{p}_{y}^{2}}{2m}+V\left(  z\right)  +U(r), \label{h0}%
\end{equation}
where $p_{x(y)}$ is the $x(y)$-component of the momentum operator. The
eigenstates are then separable as $\Psi_{nlk}(r,\varphi,z)=\psi_{n}%
(\varphi)R_{l}(r)Z_{k}(z)$ normalized as $\int\left\vert \psi_{n}%
(\varphi)\right\vert ^{2}{d}{\varphi=}$ $\int{r|R_{l}(r)|^{2}dr}=\int%
{|Z_{k}(z)|^{2}dz=}1.$ To simplify the problem further, we assume that the
confinement is strong enough such that only the lowest subbands $\left(
k=l=0\right)  $ are occupied, which makes the system effectively
one-dimensional (1D) in azimuthal direction. The eigenstates of Eq. (\ref{h0})
are
\begin{equation}
\psi_{n}(\varphi)=\frac{1}{\sqrt{2\pi}}e^{in\varphi} \label{Psi0}%
\end{equation}
with the energies $\varepsilon_{n}=\hbar^{2}n^{2}/\left(  2ma^{2}\right)
+\varepsilon_{0},$ where $\varepsilon_{0}$ is the confinement energy corresponding 
to $R_{0}(r)Z_{0}(z)$.

We wish to model the system in the presence of constant persistent current.
In the absence of SOI, the current operator along the ring is defined by its
expectation value
\begin{align}
\mathbf{j}_{n}^{\varphi}\left(  r,z\right)   &  =j_{n}^{\varphi}\left(
r,z\right)  \mathbf{e}_{\varphi}\\
&  =\mathbf{e}_{\varphi}\frac{e\hbar}{mr}\left\vert R_{0}(r)\right\vert
^{2}\left\vert Z_{0}(z)\right\vert ^{2}\operatorname{Im}\psi_{n}(\varphi
)\frac{\partial}{\partial\varphi}\psi_{n}(\varphi) \ ,
\end{align}
where we used ${v}_{\varphi}=-i\hbar/(mr)\partial/\partial\varphi$, and the
total current in the wire is
\begin{align}
I^{\varphi}  &  =\int\int dzdrj^{\varphi}\left(  r,z\right) \\
&  =-\frac{e\hbar}{m}\operatorname{Im}\psi_{n}(\varphi)\frac{\partial
}{\partial\varphi}\psi_{n}(\varphi)\int dr\frac{1}{r}\left\vert R_{0}%
(r)\right\vert ^{2}\int dz\left\vert Z_{0}(z)\right\vert ^{2}\\
&  =-\frac{e\hbar}{ma^{2}}\sum_{n}f_{n}\operatorname{Im}\psi_{n}(\varphi
)\frac{\partial}{\partial\varphi}\psi_{n}(\varphi).
\end{align}
where $e>0$, and we used $\int{dr\left\vert R_{0}(r)\right\vert ^{2}%
/r}=1/a^{2}$ assuming Gaussian $R_{0}$.\cite{Meijer} The current operator
(\ref{j2}) is diagonal in the basis of the states (\ref{Psi0}), which are therefore
also eigenfunctions of the current carrying system. The total current density
in the ring then reads:
\begin{equation}
I^{\varphi}=-\frac{e\hbar}{2\pi ma^{2}}\sum_{n}f_{n}n \ . \label{cur}%
\end{equation}
The projected Hamiltonian $\left\langle Z_{0}R_{0}\left\vert H_{0}\right\vert
Z_{0}R_{0}\right\rangle $ in the presence of the Lagrange multiplier term
$-A^{\varphi}\hat{I}^{\varphi}$ (parameterizing the vector potential as
$A^{\varphi}=2\pi\hbar n_{\lambda}/e$ where $n_{\lambda}$ is dimensionless) is
diagonal in the basis (\ref{Psi0}) with the energies
\begin{align}
\varepsilon_{n}  &  =E_{a}n^{2}-\frac{2\pi\hbar n_{\lambda}}{e}\frac{e\hbar
}{2\pi ma^{2}}n+\varepsilon_{0}\label{hc}\\
&  =E_{a}\left(  n-n_{\lambda}\right)  ^{2}+\tilde{\varepsilon}_{0} \ ,
\end{align}
where $E_{a}=\hbar^{2}/(2ma^{2})$ and $\tilde{\varepsilon}_{0}=\varepsilon
_{0}-E_{a}n_{\lambda}^{2}.$ At zero temperature $f_{n}=\Theta(\varepsilon
_{n}-\epsilon_{F}+\tilde{\varepsilon}_{0})$, where $\epsilon_{F}$ is the Fermi
energy and $\Theta$ the step function, therefore
\begin{equation}
I^{\varphi}\approx\frac{2e\hbar}{\pi ma^{2}}n_{\lambda}n_{F},
\end{equation}
where $n_{F}=\sqrt{\left(  \epsilon_{F}-\tilde{\varepsilon}_{0}\right)
/E_{a}}$. We assume that the number of electrons is constant under variation
of $n_{\lambda},$ which implies that $\tilde{\varepsilon}_{0}$ may be set to
zero. The current constraint $I^{\varphi}=I$ determines the effective vector
potential
\begin{equation}
n_{\lambda}=\frac{\pi ma^{2}}{2e\hbar n_{F}}I=\frac{\pi}{4e}\frac{\hbar}%
{E_{a}}\frac{I}{n_{F}} \,
\end{equation}
so that the spectrum (\ref{hc}) is fully determined. The current is
optimally accommodated by rigidly shifting the distribution function
proportional to the applied current.

Real magnetic field $B_{\mathrm{ext}}$ also generates persistent
currents.\cite{Buttiker} There is a
difference, however. The energies of a quantum ring in the presence of a real
magnetic flux $\Phi=\pi a^{2}B_{\mathrm{ext}}$ read
\begin{equation}
E_{n}=E_{a}\left(  n-\frac{\Phi}{\Phi_{0}}\right)^{2} \ , %
\end{equation}
where $\Phi_{0}=e/h$ is the flux quantum and we can identify $n_{\lambda}%
=\Phi/\Phi_{0}.$ The total energy in the presence of diamagnetic persistent
current is
\begin{equation}
E^{\prime}=\sum_{n_{F}^{\left(  -\right)  }}^{n_{F}^{\left(  +\right)  }}E_{n} \ , %
\end{equation}
where $n_{F}^{\left(  \pm\right)  }=\left\lfloor \pm\sqrt{2ma^{2}\left(
\epsilon_{F}-\varepsilon_{0}\right)  }/\hbar+\Phi/\Phi_{0}\right\rfloor \ $is
the largest integer smaller of equal $\sqrt{2ma^{2}\left(  \epsilon
_{F}-\varepsilon_{0}\right)  }/\hbar$. $E^{\prime}\left(  \Phi\right)  $ is
periodic, since the quantum numbers of the highest occupied states jump by
$\pm1$ when two states cross the Fermi energy. The current%
\begin{equation}
I^{\varphi\prime}=\frac{\partial}{\partial\Phi}E^{\prime}=-\Phi_{0}E_{a}%
\sum_{n_{F}^{\left(  -\right)  }}^{n_{F}^{\left(  +\right)  }}\left(
n-\frac{\Phi}{\Phi_{0}}\right)  \label{OsI}%
\end{equation}
oscillates as a function of $\Phi$ with the maximum
\begin{equation}
\left\vert I^{\varphi\prime}\right\vert _{\max}=NE_{a}\Phi_{0}=1.5\times
10^{-10}%
\operatorname{A}%
\frac{N}{1000}\frac{0.1%
\operatorname{\mu m}%
}{a^{2}},
\end{equation}
where $N=2\left(  n_{F}^{\left(  +\right)  }+n_{F}^{\left(  -\right)
}\right)  $ is the total number of electrons.

The Lagrange multiplier on the other hand contributes the additional term
$\hbar^{2}n_{\lambda}^{2}/\left(  2ma^{2}\right)  $, see Eq. (\ref{hc}), which
modifies the expressions to
\begin{align}
\frac{E_{n}}{E_{a}}  &  =\left(  n-n_{\lambda}\right)  ^{2}-n_{\lambda}%
^{2}=n\left(  n-2n_{\lambda}\right) \ , \\
I^{\varphi}  &  =\Phi_{0}\frac{\partial E}{\partial n_{\lambda}}=-E_{a}%
\Phi_{0}\sum_{-n_{F}-n_{\lambda}}^{n_{F}-n_{\lambda}}n. \label{curL}%
\end{align}
which agrees with Eq. (\ref{cur}).

\begin{figure}[t]
\centering
\includegraphics[trim=1cm 8cm 3.3cm 8cm, clip=true,width = \columnwidth]{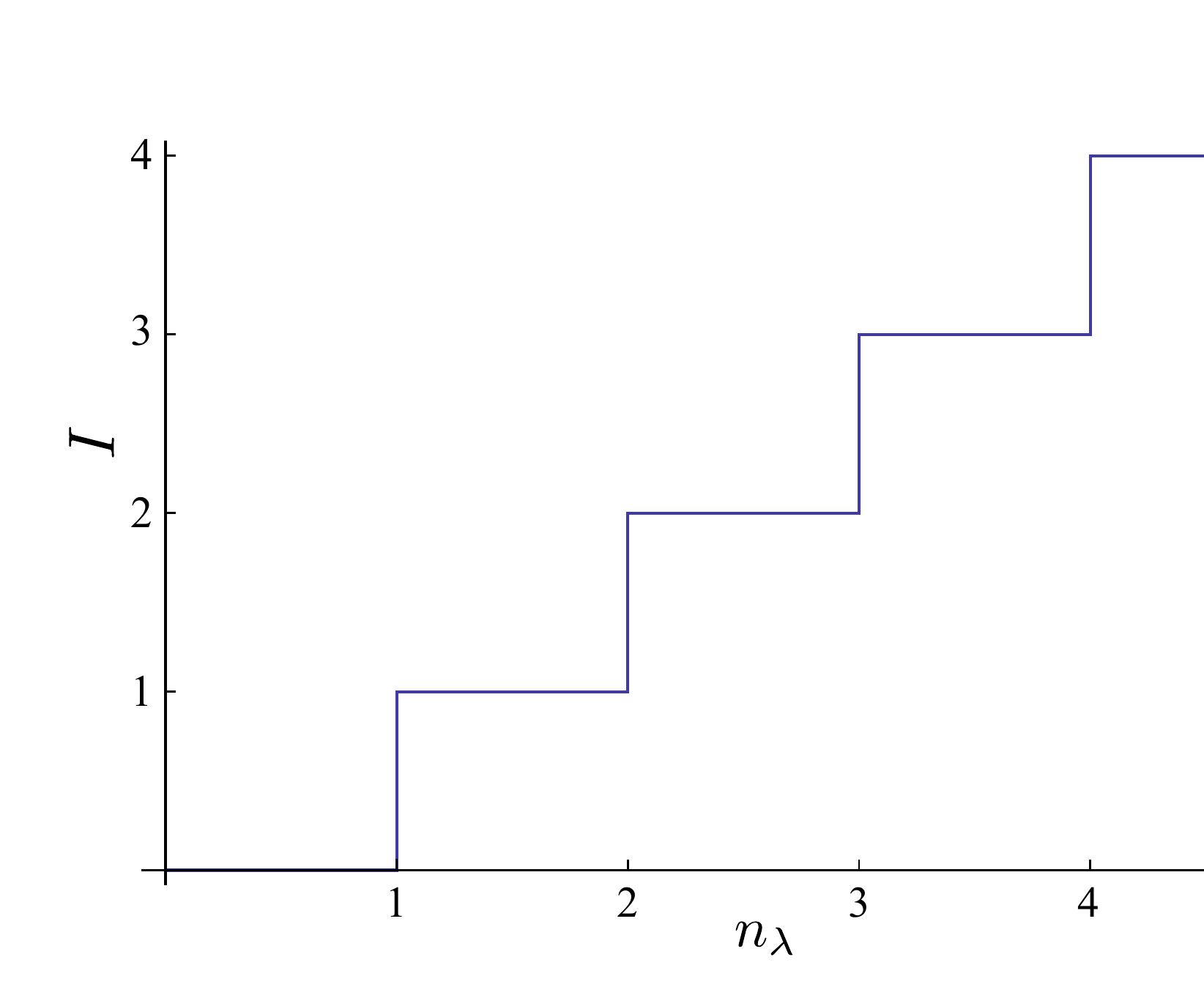}
\caption{Current versus $n_{\lambda}$. The current axis is scaled in units of
$I_{0}=2e\hbar n_{F}/(\pi ma^{2})$. }%
\label{current}%
\end{figure}

The current is finite for
any $n_{\lambda}\neq0$ (except when $N=1$ and $n=0\ $or
$2n_{\lambda}$). Thus, contrary to the diamagnetic current induced by real
magnetic field, the Lagrangian method generates unbound currents. However, due to the
discreteness of the energy levels the currents are quantized, see Fig.
\ref{current}. In the following we work with a large number of electrons such
that the currents are quasi-continuous. 

Nevertheless, if it is taken into account that there is a maximum for
magnetic-field induced currents, our method also predicts spin polarization
generated by diamagnetic
currents and correspondingly enhanced paramagnetic susceptibility of quantum
rings. It is instructive to compare the magnitudes of the
Lagrange multipliers with the corresponding magnetic fields. With
$eA^{\varphi}=2\pi n_{\lambda}=eB_{\mathrm{ext}}a^{2}/\left(  2\hbar\right)
$,
\begin{align}
B_{\mathrm{ext}}  &  =\left(  \frac{\pi\hbar}{ea}\right)  ^{2}\frac{1}{E_{a}%
}\frac{I}{n_{0}}=\left(  \frac{\pi}{e}\right)  ^{2}\frac{2m}{n_{0}%
}I\label{Bext}\\
&  \approx0.1%
\operatorname{mT}%
\frac{1000}{n_{0}}\frac{I}{\mathrm{nA}},
\end{align}
which does not depend on the size of the ring.

The generation of charge current by circularly polarized light is discussed
in Appendix \ref{appe} following Hertel.\cite{Hertel} We find that for
experimentally accessible light intensities, $B_{\mathrm{ext}}$ can be of the
order of 1 Tesla.

\section{Single mode rings in the presence of SOI \label{ring}}

In the weakly relativistic limit a particle spin experiences SOI, {\em i.e.}
effective magnetic field that scales with the particle velocity. It requires
inversion symmetry breaking induced either by space charges or asymmetric
heterostructures or by a unit cell without inversion symmetry, as is the case
for the zinkblende structure. The Rashba SOI in quasi-two-dimensional
electron gas (2DEG) is a simple realization of the former,\cite{rash1,rash2}
while the Dresselhaus\cite{dress} SOI represents the latter type. For 2DEG in
the $x,y$-plane the Hamiltonian (\ref{h0}) is then augmented by
\begin{equation}
H_{SO}^{e}=\frac{\alpha_{e}}{\hbar}\left(  \sigma_{y}p_{x}-\sigma_{x}%
p_{y}\right)  +\frac{\beta_{e}}{\hbar}\left(  \sigma_{x}p_{x}-\sigma_{y}%
p_{y}\right)  ,
\end{equation}
where $\sigma_{x(y)}$ are the $x(y)$-components of the momentum operator for
electrons and a vector of the Pauli matrices, respectively. In two-dimensional
hole gas (2DHG), on the other hand,\cite{Winkler1,Winkler2,Bulaev}
\begin{equation}
H_{SO}^{h}=\left(  i\frac{\alpha_{h}}{\hbar^{3}}{p}_{-}^{3}+\frac
{\beta_{h}}{\hbar^{3}}{p}_{-}{p}_{+}{p}_{-}\right)  \sigma
_{+}+\mathrm{h.c.},
\end{equation}
and ${O}_{\pm}={O}_{x}\pm i{O}_{y}$, where $\mathbf{{O}}%
\equiv\mathbf{{p}},\boldsymbol{\sigma}$, are the momentum operator and
the Pauli spin matrix vectors, respectively. $\alpha_{e(h)}$ and $\beta_{e(h)}$
parameterize the linear (cubic) Rashba and the linear (cubic) Dresselhaus SOI.
The canonical velocity operators are modified by the spin-orbit interaction
since they do not commute with the Hamiltonian. Dropping the index for
electrons and holes,
\begin{align}
\mathbf{ {v}}  &  =\mathbf{\dot{r}}=\frac{1}{i\hbar}\left[  \mathbf{r}%
,H\right] \label{v}\\
&  =\mathbf{ {v}}_{0}+\mathbf{ {v}}_{SO}=\frac{\hbar}{im}%
\boldsymbol{\nabla+}\frac{1}{i\hbar}\left[  \mathbf{r},H_{SO}\right]  \ .
\end{align}
where $\mathbf{ {v}}_{SO}$ is the anomalous velocity. The current operators
are modified analogously.

As before, we add an axially symmetric confinement potential to the 2DE(H)G
and consider the electric quantum confinement (1D) limit. Here,
 we separately discuss electrons and holes in such quantum rings in
the presence of a circular current, and calculate the current-induced spin polarization
in each system.

\subsection{Electrons with the Rashba SOI\label{elRash}}

For electrons in the 1D quantum ring the projection of the full Hamiltonian
$ {H}$ onto the azimuthal subspace, leads to\cite{Meijer}
\begin{align}
&H\left(  \varphi\right) =\left\langle Z_{0}R_{0}\left\vert H_{0}%
+H_{SO}\right\vert Z_{0}R_{0}\right\rangle =-\frac{\hbar^{2}}{2ma^{2}}%
\partial_{\varphi}^{2}\nonumber\\
&-i\frac{\alpha}{a}\left\{  \left(  \sigma_{x}\cos\varphi+\sigma_{y}%
\sin\varphi\right)  \partial_{\varphi}+\frac{1}{2}\left(  \sigma_{y}%
\cos\varphi-\sigma_{x}\sin\varphi\right)  \right\} \nonumber\\
&-i\frac{\beta}{a}\left\{  \left(  \sigma_{x}\sin\varphi+\sigma_{y}%
\cos\varphi\right)  \partial_{\varphi}+\frac{1}{2}\left(  \sigma_{x}%
\cos\varphi-\sigma_{y}\sin\varphi\right)  \right\}  . \label{h}%
\end{align}
Let us first focus on the Rashba spin-orbit interaction, i.e.\ $\beta=0$.
The eigenstates of the system are
\begin{align}
\psi_{n+}^{R}\left(  \varphi\right)   &  =\frac{1}{\sqrt{2\pi}}e^{in\varphi
}\left(
\begin{array}
[c]{c}%
\cos\frac{\theta_{R}}{2}\\
\sin\frac{\theta_{R}}{2}e^{i\varphi}%
\end{array}
\right)  ;\label{psiLR1}\\
\psi_{n-}^{R}\left(  \varphi\right)   &  =\frac{1}{\sqrt{2\pi}}e^{in\varphi
}\left(
\begin{array}
[c]{c}%
-\sin\frac{\theta_{R}}{2}\\
\cos\frac{\theta_{R}}{2}e^{i\varphi}%
\end{array}
\right)  , \label{psiLR2}%
\end{align}
where $n$ is an integer, with the energies%
\begin{equation}
\frac{E_{n\sigma}}{E_{a}}=\left(  n+\frac{1}{2}\right)  ^{2}+\sigma\left(
n+\frac{1}{2}\right)  \sec\theta_{R}+\frac{1}{4},
\end{equation}
where $\tan\theta_{R}=2ma\alpha/\hbar^{2}$. The velocity operator in this
system reads
\begin{equation}
 {v}_{\varphi}=-\frac{i\hbar}{ma}\partial_{\varphi}+\frac{\alpha}{\hbar
}\sigma_{r}.
\end{equation}
and current is $\left\langle {{I}}^{\varphi}\right\rangle =\sum_{n\sigma
}f_{n\sigma}{{I}}_{n\sigma}^{\varphi}=I$. The current operator is diagonal in
the $n\sigma$ basis (Eqs. (\ref{psiLR1}-\ref{psiLR2})), but acquires a
spin dependence,%
\begin{align}
{{I}}_{n\sigma}^{\varphi}  &  =-\frac{e\hbar}{2\pi ma^{2}}n-\sigma
\frac{e\alpha}{2\pi\hbar a}\sin\theta\nonumber\\
&  -\frac{e\hbar}{2\pi ma^{2}}\left(  \delta_{\sigma,+1}\cos^{2}\frac{\theta
}{2}+\delta_{\sigma,-1}\sin^{2}\frac{\theta}{2}\right) \ . \label{Inse}%
\end{align}
The projected Hamiltonian in the presence of the Lagrange multiplier term
(parameterizing the vector potential as $A^{\varphi}=\hbar n_{\lambda}/e$
where $n_{\lambda}$ is dimensionless) is diagonal in the basis
(\ref{Psi0}) with the energies%
\begin{align}
\frac{E_{n\sigma}}{E_{a}}  &  =\left(  n+\frac{1}{2}\right)  ^{2}%
+\sigma\left(  n+1/2\right)  \sec\theta_{R}+\frac{1}{4}\nonumber\\
&  -\frac{2\pi\hbar n_{\lambda}}{e}\frac{2ma^{2}}{\hbar^{2}}\left[
\frac{e\hbar}{2\pi ma^{2}}n+\sigma\frac{e\alpha}{2\pi\hbar a}\sin\theta
_{R}\right. \nonumber\\
&  \left.  +\frac{e\hbar}{2\pi ma^{2}}\left(  \delta_{\sigma,+1}\cos^{2}%
\frac{\theta_{R}}{2}+\delta_{\sigma,-1}\sin^{2}\frac{\theta_{R}}{2}\right)
\right] \\
&  =\left(  n-n_{\lambda}+\frac{1}{2}\right)  ^{2}+\sigma\left(  n-n_{\lambda
}+\frac{1}{2}\right)  \sec\theta_{R}+\frac{1}{4}-n_{\lambda}^{2}.%
\end{align}
At zero temperature:
\begin{equation}
I=\sum_{n\sigma}f_{n\sigma}{{I}}_{n\sigma}^{\varphi}=\sum_{\sigma}\sum
_{-n_{r}+n_{\lambda\lambda}-\sigma\sec\theta_{R}-1/2}^{n_{r}+n_{\lambda
\lambda}-\sigma\sec\theta_{R}-1/2}I_{n\sigma}^{\varphi}=\frac{2e\hbar}{\pi
ma^{2}}n_{\lambda}n_{r},%
\end{equation}
where $n_{r}\equiv\sqrt{\epsilon_{F}/E_{a}+  {\sec}^{2}{\theta}%
_{R} /4}$ and we substituted Eq. (\ref{Inse}). The leading term is
therefore the same as in the absence of spin-orbit interaction:
\begin{equation}
n_{\lambda}=\frac{\pi}{4e}\frac{\hbar}{E_{a}}\frac{I}{n_{r}}. \label{nl}%
\end{equation}

Since the system is not magnetic, it is not spin polarized at
equilibrium.
The spin polarization of the current-carrying ground
state reads
\[
\langle\sigma_{z}\rangle_{I}^{R}=\sum_{n\sigma}\left\langle \psi_{n\sigma}%
^{R}\left\vert  {\sigma}_{z}\right\vert \psi_{n\sigma}^{R}\right\rangle
_{I}=\sum_{n\sigma}f_{n\sigma}\sigma\cos\theta_{R} \ ,
\]
and $\langle\sigma_{y}\rangle_{I}^{R}=\langle\sigma_{x}\rangle_{I}^{R}=0$. In
the absence of current the energy bands are equally filled for both spins in
the negative and positive directions, and we do not have unpaired electrons.
Thus:
\begin{align}
\langle\sigma_{z}\rangle_{I=0}^{R}  &  =\cos\theta_{R}\sum_{n\sigma}%
\sigma\Theta(\epsilon_{F}-E_{n\sigma})\\
&  =\cos\theta_{R}\sum_{\sigma}\sum_{-n_{r}+n_{\lambda}-\sigma\sec\theta
_{R}-1/2}^{n_{r}+n_{\lambda}-\sigma\sec\theta_{R}-1/2}\sigma=0.
\end{align}
In the presence of the current bias the electron distribution is shifted in
reciprocal space around the Fermi level by $n_{\lambda}$. The
spinors Eqs. (\ref{psiLR1},\ref{psiLR2}) that determine the spin texture do
not depend on $n.$ Furthermore, the relative occupation of the two spin bands
also remains the same. Therefore, the induced current does not generate spin
polarization and $\langle\sigma_{z}\rangle_{I}^{R}=0$ for all current levels.
To put it differently, since the ring is invariant to rotation, the system is
invariant to a Galilean gauge transformation that induces the persistent
current. The conclusion that there is no current-induced spin accumulation
in the Rashba systems holds also for 1D wires. Vanishing of the spin
accumulation is caused by the compensating effect of the two subbands. This
can be suppressed when a gap is induced at $k=0$ by a Zeeman field or exchange
interfaction and the Fermi energy is tuned to fall into this
gap.\cite{Joibari} We also note that the linear current-induced spin accumulation
does not vanish in two-dimensional electron gas either.\cite{Edelstein}

\subsection{Electrons with the Dresselhaus SOI\label{elDress}}

A similar situation arises for a ring with only linear Dresselhaus
interaction, {\em i.e.} \ $\alpha=0$ in Eq. (\ref{h}). Its eigenstates
are\cite{ring}
\begin{align}
\psi_{n+}^{D}\left(  \varphi\right)   &  =\frac{1}{\sqrt{2\pi}}e^{in\varphi
}\left(
\begin{array}
[c]{c}%
-\sin\frac{\theta_{D}}{2}\\
i\cos\frac{\theta_{D}}{2}e^{-i\varphi}%
\end{array}
\right)  ;\label{psiLD1}\\
\psi_{n-}^{D}\left(  \varphi\right)   &  =\frac{1}{\sqrt{2\pi}}e^{in\varphi
}\left(
\begin{array}
[c]{c}%
\cos\frac{\theta_{D}}{2}\\
i\sin\frac{\theta_{D}}{2}e^{-i\varphi}%
\end{array}
\right)  , \label{psiLD2}%
\end{align}
with the energies identical to those for the Rashba
ring%
\begin{equation}
\frac{E_{n\sigma}}{E_{a}}=\left(  n+\frac{1}{2}\right)  ^{2}+\sigma\left(
n+\frac{1}{2}\right)  \sec\theta_{D}+\frac{1}{4},
\end{equation}
but now $\tan\theta_{D}=2ma\beta/\hbar^{2}$. Thus, the spin texture does not
depend on the angular momentum. This means that shifting a distribution
function rigidly does not change the balance of the spin states,
and as in the Rashba case, there is no current-induced spin polarization.

\subsection{Holes with the Dresselhaus SOI\label{holesDress}}

\begin{figure}[t]
\centering
\includegraphics[trim=0cm 10cm 1cm 8cm, clip=true,width = \columnwidth]{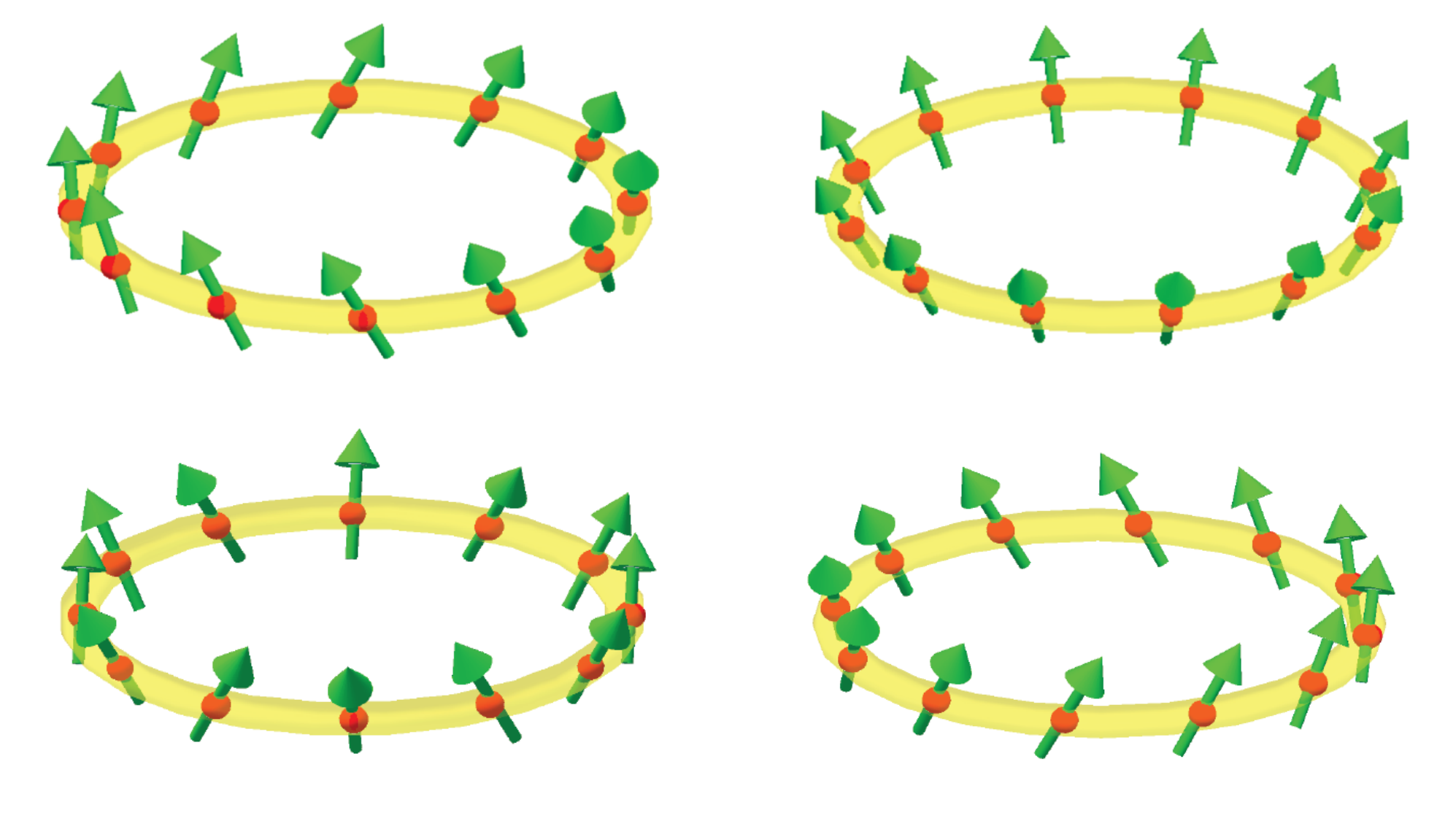} \caption{Top:
 Electrons spin texture in the presence of the linear Dresselhaus (left) or
Rashba (right) interaction. The angle of the spin with the ring in this case only depends to
the SO coupling strength and not the angular momentum. \newline
Bottom:
The hole spin
texture in the presence of the cubic Dresselhaus (left) or Rashba (right)
spin-orbit interaction. The direction of the spins in both cases depends on
the angular velocity of the holes and on the SO coupling strength.\newline
Here, the particles orbit counterclockwise. The clockwise
movement induces spin texture which is mirrored with respect to the plane
containing the ring.}%
\label{fig1}%
\end{figure}

Stepanenko \textit{et al.}\cite{stepa} derived an effective low energy
Hamiltonian for heavy holes from the Luttinger Hamiltonian that includes
Dresselhaus and Rashba like SOI that are cubic in the angular momenta. Simple
analytical solutions were obtained in two limits, representing the
Dresselhaus-only interaction ($\alpha_{h}=0$) and the Rashba-only SOI
($\beta_{h}=0$). The spin textures in these two limits are shown in Fig.
\ref{fig1}. In contrast to the electron case, both spinors include terms
quadratic in the angular momentum. Here, we show that these do generate
current-induced spin accumulation. 

The heavy-hole Hamiltonian for a 1D ring with the Dresselhaus SOI is\cite{stepa}
\begin{align}
{H}_{0}^{cD}  &  =-\frac{\hbar^{2}}{2m_{hh}a^{2}}\partial_{\varphi}%
^{2}\nonumber\\
&  +\beta_{h}e^{i\varphi}\left(  G_{0}+G_{1}\partial_{\varphi}+G_{2}%
\partial_{\varphi}^{2}+G_{3}\partial_{\varphi}^{3}\right)  \sigma
_{-}\nonumber\\
&  +\beta_{h}e^{-i\varphi}\left(  G_{0}-G_{1}\partial_{\varphi}+G_{2}%
\partial_{\varphi}^{2}-G_{3}\partial_{\varphi}^{3}\right)  \sigma_{+}\ ,
\label{h0d}%
\end{align}
where $G_{0}=i(R_{0}+R_{1}-R_{2})$, $G_{1}=-(R_{1}+R_{2})$, $G_{2}%
=i(R_{2}-2R_{3})$, and $G_{3}=-R_{3}$, and the coefficients $R_{j}=\left\langle
r^{-j}\partial_{r}^{3-j}\right\rangle _{\mathrm{radial}}$ depend on the ground
state radial confinement wave function. For a ring with radius $a$ and width
$w,$ $R_{2}=R_{3}/2=1/(2a^{3})$ and $R_{1}=-2/3R_{0}=-1/(aw^{2})$%
.\cite{Kovalev} Here, $m_{hh}=m_{0}/(\gamma_{1}+\tilde{\gamma})$, where
$\tilde{\gamma}=\gamma_{2}$ for the $[001]$ ($\tilde{\gamma}=\gamma_{3}$ for
$[111]$) growth direction and $\gamma_{i}$ are the standard Luttinger
parameters for the valence band of \textrm{III}-\textrm{V} semiconductors. The
eigenfunctions of the system are
\begin{align}
\psi_{l,+}^{cD}  &  =\frac{1}{\sqrt{2\pi}}e^{il\varphi}\left(
\begin{array}
[c]{c}%
i\cos\frac{\theta^{cD}(l)}{2}e^{-i\varphi/2}\\
-\sin\frac{\theta^{cD}(l)}{2}e^{i\varphi/2}%
\end{array}
\right)  ;\label{psiCD1}\\
\psi_{l,-}^{cD}  &  =\frac{1}{\sqrt{2\pi}}e^{il\varphi}\left(
\begin{array}
[c]{c}%
i\sin\frac{\theta^{cD}(l)}{2}e^{-i\varphi/2}\\
\cos\frac{\theta^{cD}(l)}{2}e^{i\varphi/2}%
\end{array}
\right)  , \label{psiCD2}%
\end{align}
where $l=n+1/2$, and the texture angle $\theta^{cD}(l)$ is
\begin{equation}
\theta^{cD}(l)=\tan^{-1}\left[  \frac{2m_{hh}\beta_{h}}{\hbar^{2}R_{3}^{2/3}%
}\left(  \frac{2}{3}R_{0}+\left(  l^{2}-\frac{5}{4}\right)  R_{3}\right)
\right]  , \label{encD}%
\end{equation}
with the energies
\[E_{l\sigma}=E_{a}^{h}\left(  l^{2}+\frac{1}{4}+\sigma l\sec\theta^{cD}\left(l\right)\right),\]
where $E_{a}^{h}=\hbar^{2}/(2m_{hh}a^{2})$. In terms of the velocity operator
\begin{align}
{ {v}}_{\varphi}  &  =-\frac{i\hbar}{m_{hh}a}\partial_{\varphi}%
+\frac{ia\beta_{h}}{\hbar}e^{i\varphi}\left(  G_{1}+2G_{2}\partial_{\varphi
}+3G_{3}\partial_{\varphi}^{2}\right)  \sigma_{-}\nonumber\\
&  +\frac{ia\beta_{h}}{\hbar}e^{-i\varphi}\left(  -G_{1}+2G_{2}\partial
_{\varphi}-3G_{3}\partial_{\varphi}^{2}\right)  \sigma_{+},
\end{align}
the current operator reads
\begin{equation}
\hat{I}_{l\sigma\sigma^{\prime}}^{\varphi}=\frac{e}{a}\operatorname{Re}%
\psi_{l\sigma}^{\dag}(\varphi){ {v}}_{\varphi}\psi_{l\sigma^{\prime}%
}(\varphi).
\end{equation}

Both the Hamiltonian and current operators are diagonal in the orbital angular
momentum, which allows us to introduce $2\times2$ operators in spin space
for calculation of the expectation values in position space:\textit{ }
\begin{align}
-\frac{\hat{I}_{l}^{\varphi}}{e}=  &  2E_{a}^{h}\left(  l_{-} {\sigma}%
_{1}+l_{+} {\sigma}_{2}\right) \nonumber\\
&  +\frac{\beta_{h}}{\hbar}\left(  G_{1}+2iG_{2}l_{-}+3G_{3}l_{-}^{2}\right)
\sigma_{-}\nonumber\\
&  +\frac{\beta_{h}}{\hbar}\left(  G_{1}-2iG_{2}l_{+}+3G_{3}l_{+}^{2}\right)
\sigma_{+},
\end{align}
where $ {\sigma}_{1}$, and $ {\sigma}_{2}$ are $2\times2$ matrices with
all elements zero except for the first and second diagonal one, respectively,
and $l_{\pm}=l\pm1/2$. Thus, the Hamiltonian ${H}^{cD}+\lambda\left\langle
{{I}}^{\varphi}\right\rangle $ in spin space reads
\begin{align}
{H}_{l}^{cD}  &  =E_{a}^{h}\left[  \left(  l_{-}^{2}-n_{\lambda}l_{-}\right)
 {\sigma}_{1}+\left(  l_{+}^{2}-n_{\lambda}l^{+}\right)   {\sigma}%
_{2}\right] \nonumber\\
&  -\beta_{h}\left(  G_{0}+G_{1}l_{-}+G_{2}l_{-}^{2}+G_{3}l_{-}^{3}\right)
\sigma_{-}\nonumber\\
&  -\beta_{h}\left(  G_{0}-G_{1}l_{+}+G_{2}l_{+}^{2}-G_{3}l_{+}^{3}\right)
\sigma_{+}\nonumber\\
&  +n_{\lambda}\beta_{h}\left(  G_{1}+2G_{2}l_{-}+3G_{3}l_{-}^{2}\right)
\sigma_{-}\nonumber\\
&  +n_{\lambda}\beta_{h}\left(  -G_{1}+2G_{2}l_{+}-3G_{3}l_{+}^{2}\right)
\sigma_{+}\ .
\end{align}
The eigenstates in the presence of a current now read
\begin{align}
\psi_{(l,n_{\lambda}),+}^{cD}  &  =\frac{1}{\sqrt{2\pi}}e^{il\varphi}\left(
\begin{array}
[c]{c}%
\cos\frac{\theta^{cD}(l,n_{\lambda})}{2}e^{-(i/2)(\varphi+\pi/2)}\\
\sin\frac{\theta^{cD}(l,n_{\lambda})}{2}e^{(i/2)(\varphi+\pi/2)}%
\end{array}
\right)  ;\\
\psi_{(l,n_{\lambda}),-}^{cD}  &  =\frac{1}{\sqrt{2\pi}}e^{il\varphi}\left(
\begin{array}
[c]{c}%
-\sin\frac{\theta^{cD}(l,n_{\lambda})}{2}e^{-(i/2)(\varphi+\pi/2)}\\
\cos\frac{\theta^{cD}(l,n_{\lambda})}{2}e^{(i/2)(\varphi+\pi/2)}%
\end{array}
\right)  ,
\end{align}
with the spin texture
\begin{align}
\theta^{cD}(l,n_{\lambda})  &  =\arctan\\
&  \left[  \frac{2m_{hh}\beta_{h}}{\hbar^{2}R_{3}^{2/3}}\left(  \frac{2}%
{3}R_{0}+\left(  l^{2}-\frac{5}{4}-3n_{\lambda}^{2}+2\frac{n_{\lambda}^{3}}%
{l}\right)  R_{3}\right)  \right]  \ , \nonumber\label{bn}%
\end{align}
and the energies
\[
E_{l\sigma}=E_{a}\left(  \left(  l-n_{\lambda}\right)  ^{2}+\frac{1}{4}%
+\sigma\frac{\left(  l-n_{\lambda}\right)  }{\cos\theta^{cD}\left(  \left(
l-n_{\lambda}\right)  ,n_{\lambda}\right)  }\right)  .
\]
We can obtain $n_{\lambda}$ from the current constraint by noting that the
state $l\sigma$ carries the current
\begin{align}
{{I}}_{l\sigma}^{\varphi}  &  =-\frac{e}{2\pi}\left\{  \frac{\hbar}{ma^{2}%
}l-\sigma\frac{\hbar}{2ma^{2}}\cos\theta^{cD}(l)\right. \nonumber\\
&  \left.  +\sigma\frac{\beta_{h}}{\hbar}\left(  G_{1}-iG_{2}-3G_{3}\left[
l^{2}+\frac{1}{4}\right]  \right)  \sin\theta^{cD}(l)\right\}  .
\end{align}

We now derive analytical expressions for $n_{\lambda}$ in the weak spin-orbit
coupling limit, {\em i.e.} for small $\theta^{cD}$. Subsequently, we also present
numerical results for larger SOI strengths. For small angles $\theta^{cD}$,
the expectation value of the current reduces to
\begin{align}
{{I}}_{l\sigma}^{\varphi} &  \approx-\frac{e}{2\pi}\left\{  \frac{\hbar
}{ma^{2}}l-\sigma\frac{\hbar}{2ma^{2}}\right.  \nonumber\\
&  +\sigma\frac{\beta_{h}^{2}}{\hbar E_{a}^{h}}\left(  G_{1}-iG_{2}%
-3G_{3}\left[  l^{2}+\frac{1}{4}\right]  \right)  \times\nonumber\\
&  \left.  \left(  \frac{2}{3}R_{0}+\left(  l^{2}-\frac{5}{4}-3n_{\lambda}%
^{2}+2\frac{n_{\lambda}^{3}}{l}\right)  R_{3}\right)  \right\}.
\end{align}
At zero temperature%
\begin{equation}
I_{\varphi}=\sum_{\sigma=\pm1}\sum_{-n_{r}+n_{\lambda}-\sigma/2}%
^{n_{r}+n_{\lambda}-\sigma/2}{{I}}_{l\sigma}^{\varphi}=I,
\end{equation}
where $n_{r}\approx\sqrt{\epsilon_{F}/E_{a}^{h}}$ and $\epsilon_{F}$ is the Fermi energy in
the absence of current. Taking $\cos\theta^{cD}\approx1$ in the boundaries of the
summation,
\begin{equation}
n_{\lambda}\approx\frac{\hbar}{4E_{a}^{h}}\frac{\pi}{e}\frac{I}{n_{r}}\left(
1+3\left(  \frac{\beta_{h}R_{3}n_{r}}{E_{a}}\right)  ^{2}\right)  \approx
\frac{\pi\hbar}{4eE_{a}^{h}}\frac{I}{n_{r}}.
\end{equation}
which in the limit of weak SOI does not depend on $\beta_{h}$. We find that
the system is now spin-polarized in the $z$-direction. With
\begin{align}
\left\langle \mathbf{\sigma}_{z}\right\rangle _{l\sigma}&=\langle
\psi_{(l,n_{\lambda})\sigma}^{cD}|\sigma_{z}|\psi_{(l,n_{\lambda})\sigma}%
^{cD}\rangle \nonumber\\
&=\sigma\cos\theta^{cD}(l,n_{\lambda
})\approx\sigma\left(  1-\theta^{cD}%
(l,n_{\lambda})^{2}\right)  \ ,
\end{align}
the total spin polarization is
\begin{align}
\left\langle {\sigma}_{z}\right\rangle _{I}^{cD} &  =\sum_{n\sigma}\sigma
f_{n\sigma}\cos\theta^{cD}(l,n_{\lambda
})\nonumber\label{sz1}\\
&  \approx\sum_{\sigma=\pm1}\sum_{n=-n_{r}-n_{\lambda}+\sigma/2}%
^{n_{r}-n_{\lambda}+\sigma/2}\sigma\left(  1-\left(  \theta^{cD}(l,n_{\lambda
})\right)  ^{2}\right).
\end{align}
This leads to
\begin{align}
\left\langle \mathbf{\sigma}_{z}\right\rangle _{I}^{cD} &  =\frac{\beta
_{h}^{2}}{\left(  E_{a}^{h}\right)  ^{2}a^{6}}\left\{  -4n_{\lambda}^{3}%
n_{r}+64\frac{n_{\lambda}^{6}}{n_{r}^{3}}\left(  n_{\lambda}-\frac{1}%
{2}\right)  \right.  \nonumber\\
&  +2n_{r}n_{\lambda}\left(  4n_{\lambda}^{2}+1+\left(  n_{r}+1\right)
\left(  2n_{r}+1\right)  \right)  \nonumber\\
&  \left.  -\left(  \frac{a^{2}}{w^{2}}-\left(  \frac{5}{4}+3n_{\lambda}%
^{2}\right)  \right)  \left(  -4n_{\lambda}n_{r}-16\frac{n_{\lambda}^{4}%
}{n_{r}^{2}}\right)  \right\}  .
\end{align}
To the leading order in the current $n_{\lambda}$:%
\begin{align}
\left\langle \mathbf{\sigma}_{z}\right\rangle _{I}^{cD} &  \rightarrow
\frac{2\beta_{h}^{2}}{\left(  E_{a}^{h}\right)  ^{2}}\frac{n_{r}n_{\lambda}%
}{a^{6}}\nonumber\\
&  \left[  1+\left(  n_{r}+1\right)  \left(  2n_{r}+1\right)  +\frac{2a^{2}%
}{w^{2}}-\frac{5}{2}\right].
\end{align}
The total number of electrons
\begin{equation}
N=\sum_{n\sigma}f_{n\sigma}\approx\sum_{\sigma=\pm1}\sum_{-n_{r}+n_{\lambda
}-\sigma/2}^{n_{r}+n_{\lambda}-\sigma/2}1=4n_{r}.%
\end{equation}
For $n_{r}\gg a/w$ the term proportional to $2n_{r}^{3}$ dominates and the
spin polarization simplifies to
\begin{equation}
\left\langle \mathbf{\sigma}_{z}\right\rangle _{I_{\varphi}}^{cD}\approx
\frac{4\beta_{h}^{2}}{E_{a}^{h}{}^{2}}R_{3}^{2}n_{r}^{3}n_{\lambda}%
=\frac{\epsilon_{F}}{E_{a}^{h}}\frac{\pi\hbar\beta_{h}^{2}}{\left(  E_{a}%
^{h}a^{2}\right)  ^{3}}\frac{I}{e},\label{SZF}%
\end{equation}
while in the limit of a wide and narrow ring
\begin{equation}
\left\langle \mathbf{\sigma}_{z}\right\rangle _{I_{\varphi}}^{cD}%
\approx\left\langle \mathbf{\sigma}_{z}\right\rangle _{I}^{cD}\rightarrow
\frac{4\beta_{h}^{2}}{\left(  E_{a}^{h}a^{2}\right)  ^{2}}\frac{n_{r}%
n_{\lambda}}{w^{2}}=\frac{a^{2}}{w^{2}}\frac{\pi\hbar\beta_{h}^{2}}{\left(
E_{a}^{h}a^{2}\right)  ^{3}}\frac{I}{e}.
\end{equation}
The spin polarization is in both cases proportional to the current and the
squared amplitude of the SOI interaction, which is expected. The
proportionality with Fermi energy when $n_{r}\gg a/w$ reflects the increasing
spin texture angle $\theta^{cD}$ with energy. This implies the scaling with the
squared number of particles as well as the area of the ring. In the opposite
limit, we find that the spin polarization increases when tightening the
laterally quantized subband, because this enhances the SOI matrix elements.
For realistic and currently experimentally feasible dimensions, the former
approximation seems more appropriate, and thus, we focus on this limit
henceforth. One can estimate the the spin polarization in this regime from Eq.
(\ref{SZF}) and the Dresselhaus coupling constant for GaAs\cite{SO} $\beta
_{h}=30\text{ }\mathrm{eV\mathring{A}}^{3}$ as:
\begin{equation}
\left\langle \mathbf{\sigma}_{z}\right\rangle _{I_{\varphi}}^{cD}%
\approx0.2\left(  \frac{\epsilon_{F}}{10\,\text{\textrm{meV}}}\right)  \left(
\frac{a^{2}}{\mathrm{\mu}\text{\textrm{m}}}\right)  \left(  \frac
{I}{\mathrm{n}\text{\textrm{A}}}\right)  \left(  \frac{\beta_{h}%
}{30\,\text{\textrm{eV}}\mathrm{\mathring{A}}^{3}}\right)  ^{2}.%
\end{equation}

For better understanding we can derive the equivalent effective magnetic
field that would generate the same spin polarization (\ref{SZF}) in the
absence of SOI.
Consider the Hamiltonian $H_{B}=p^{2}/(2m_{hh})\hat{1}-\Delta\sigma_{z}$, with
the Zeeman energy $\Delta=\hbar eg_{h}B_{eff}/(4m_{hh})$, where $g_{h}$ is the
gyromagnetic ratio. Clearly such a system is spin polarized and in the limit
of $\Delta/\epsilon_{F}\ll1$:
\begin{equation}
\left\langle \sigma_{z}\right\rangle _{Z}\approx\frac{\hbar}{2}\frac
{eg_{h}B_{eff}}{m_{hh}}\frac{1}{\sqrt{E_{a}^{h}\epsilon_{F}}}.\label{SB}%
\end{equation}
The $\epsilon_{F}^{-1/2}$ dependence reflects the 1D density of states that
decreases with energy. Comparison of Eqs. (\ref{SB}) and (\ref{SZF}) gives an
equivalent effective field of \thinspace%
\[
B_{eff}=\frac{32\pi}{\sqrt{2}}\frac{m_{hh}^{9/2}a}{e\hbar^{7}g_{h}}%
\epsilon_{F}^{3/2}\beta_{h}^{2}\frac{I}{e},%
\]
where we assume the g-factor $g_{h}=-0.5$.\cite{g}
\begin{figure}[t]
\centering
\includegraphics[trim=1cm 7cm 4.5cm 0.3cm, clip=true,width = \columnwidth]{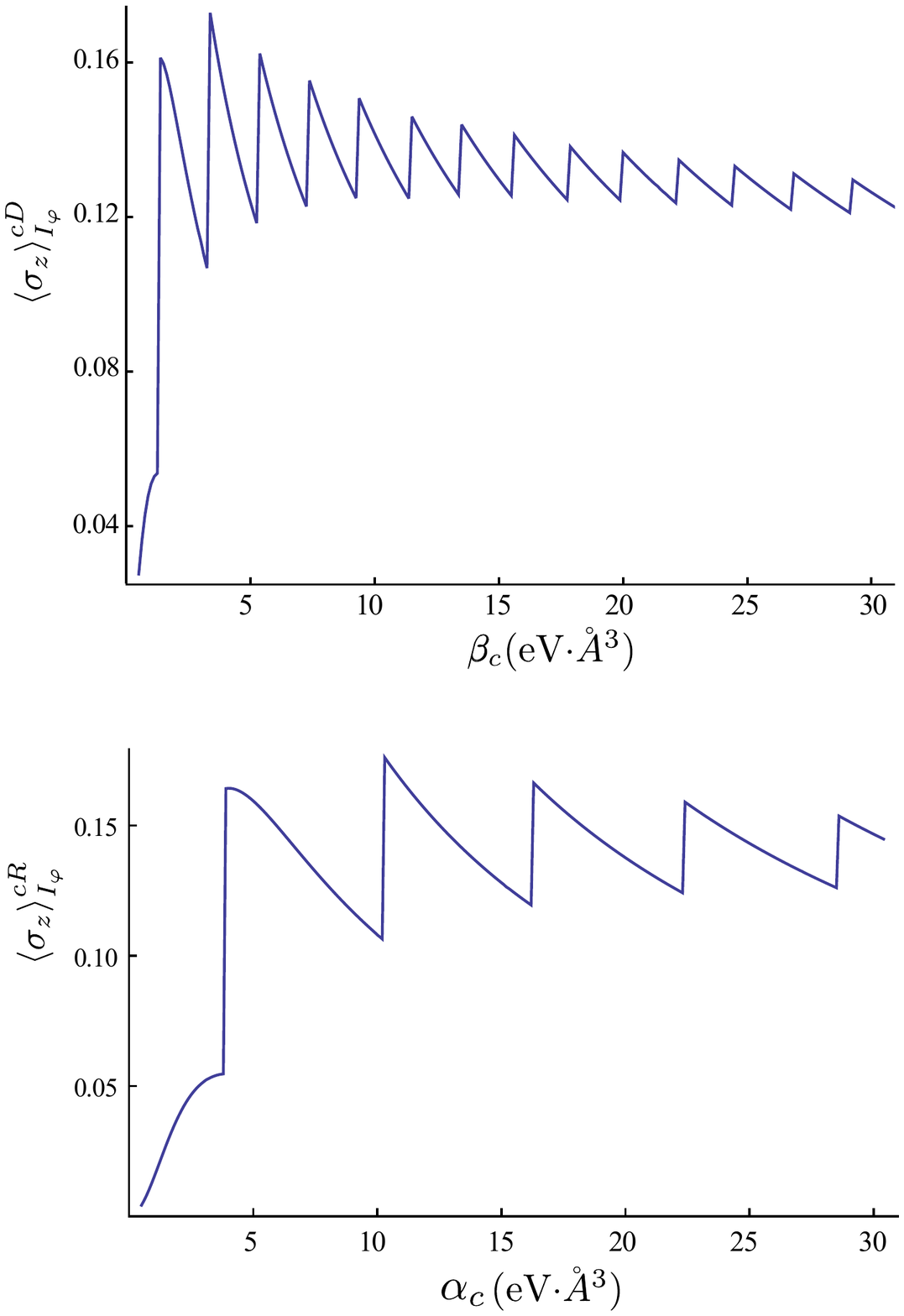}
\caption{The current-induced spin polarization of heavy holes in a quantum
ring subject to the cubic Dresselhaus (Rashba) SOI, plotted in the upper
(lower) panel. The plots are shown as a function of the SOI parameters and
$m_{hh}=0.45m_{0}$, $N=1144,$ which in the absence of current is equivalent to
$\epsilon_{F}=10\;\text{meV}$, a radius of $a=1$ $\mathrm{\mu}$m, and width of
$w=50$ nm. Here, we assumed a current of $I=35$ nA, which is equivalent to
circularly polarized light with the frequency of $\omega=2\times10^{14}%
$s$^{-1}$, and the electric field amplitude of $|E_{0}|=\sqrt{60}\times10^{7}%
$Vm$^{-1}$, see appendix \ref{appe}. }%
\label{numeric}%
\end{figure}
Inserting
the parameters
\begin{equation}
B_{eff}=1.3\left(  \frac{\epsilon_{F}}{10\,\mathrm{meV}}\right)  ^{3/2}%
\frac{a}{1\,\mathrm{\mu m}}\frac{I}{\mathrm{nA}}\left(  \frac{\beta_{h}%
}{30\,\mathrm{eV\mathring{A}}^{3}}\right)  ^{2}\mathrm{mT} \ ,%
\end{equation}
we find that light-induced current of the order of 10 nA (see Appendix)
generates the effective field of roughly 10 mT.

\begin{figure}[h]
\centering
\includegraphics[trim=1.5cm 6cm 4.5cm 2cm, clip=true,width = \columnwidth]{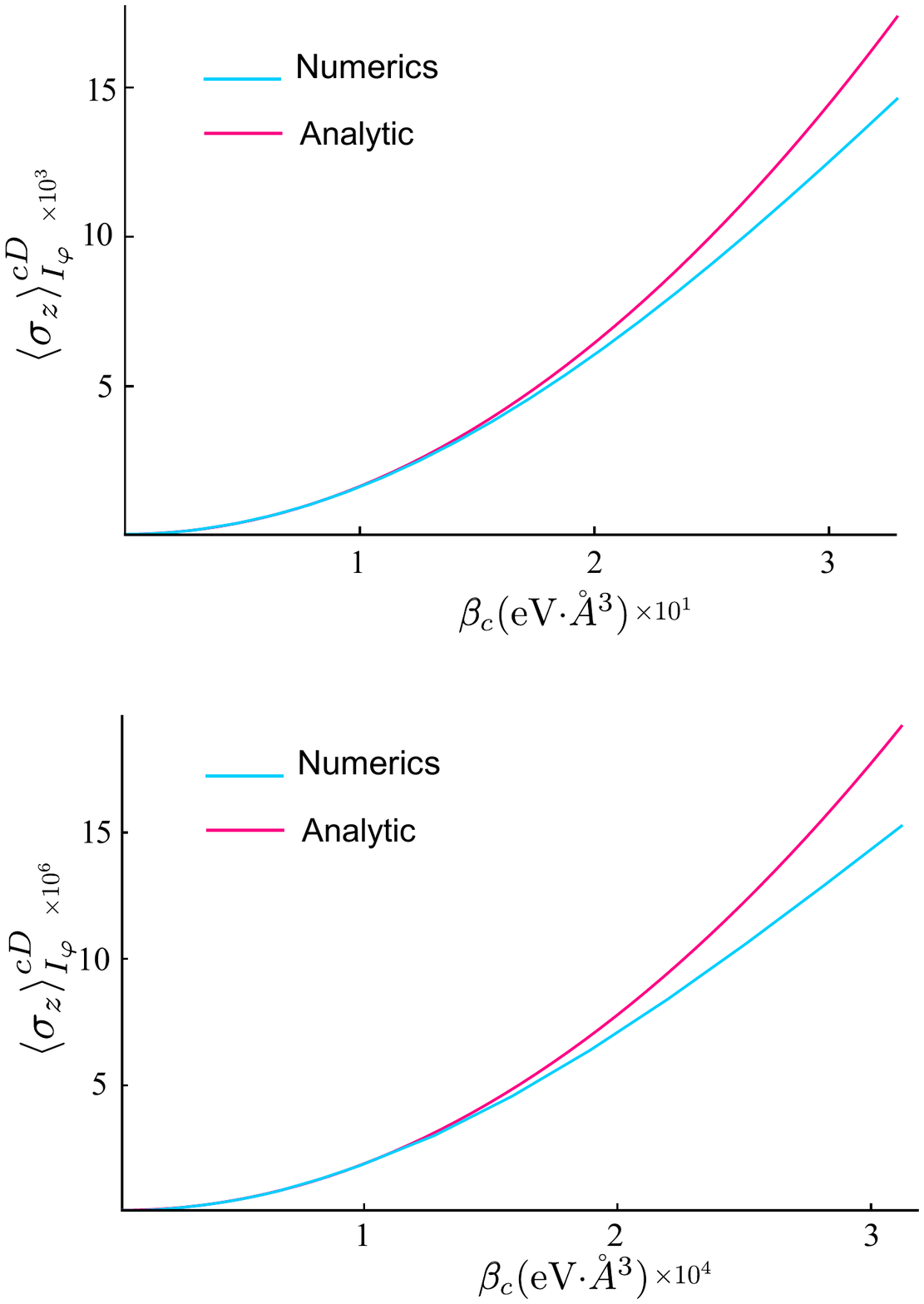}\caption{Comparison
of analytical and numerical results in the regime of the small Dresselhaus SOI for holes orbiting in a ring.
 Upper
panel: $n_{r}^{2}\gg a^{2}/w^{2}$. The parameters are the same as in Fig.
\ref{numeric} in this limit. Lower panel: Limit of $n_{r}^{2}\ll a^{2}/w^{2}$
with $w=1$ nm. The other parameters are the same as in Fig. \ref{numeric}. }%
\label{compare}%
\end{figure}

Keeping in mind that the current is quantized in steps as a function of the
system parameters as discussed above, the spin polarization computed
numerically increases linearily with the current level up to $I=100\,$%
{\normalsize nA}, in agreement with the analytic result. The deviations from
the perturbation theory are quite large for the spin-orbit interaction parameter
for GaAs used above. The non-perturbative numerical results for the spin
polarization are plotted as a function of SOI strengths for constant electron
numbers in Fig.\ \ref{numeric}. We observe that at small $\beta_{h}$
the spin polarization increases quadratically with SOI as found in the weak
SOI limit above but saturates at higher values. We also observe a saw-tooth
like behavior on top of this trend that is caused by a repopulation of states:
SOI induces a spin polarization when the current bias shifts the
occupation numbers around the Fermi level. For small but increasing $\beta
_{h}$ we expect an increasing spin polarization with SOI at constant current
since the state dependence of the spin texture increases. At large $\beta_{h}%
$, on the other hand, the angle of the spin with respect to the $z$%
-axis\ $\theta^{cD}(l,n_{\lambda})$ can be large, corresponding to smaller
values of the $z$-component of the spin. Thus, by further increasing
$\beta_{h}$, the overall polarization saturates and even slightly decreases.
The jumps reflect level crossings with increasing $\beta_{h}.$ In these
calculations the number of electrons is kept constant (the Fermi energy
oscillates). At such a discontinuity an electron vacates a high angular
momentum state in favor of a smaller one, which reduces the total spin
polarization. In the regime of small SOI, the numerical and analytical results
for the current induced spin polarization agree well, see Fig. \ref{compare}.

\subsection{Holes with the Rashba SOI\label{holesRash}}

The Hamiltonian of holes in the presence of the
Dresselhaus SOI is:\cite{stepa}
\begin{align}
{H}_{0}^{cR}=  &  -\frac{\hbar^{2}}{2m_{hh}a^{2}}\partial_{\varphi}%
^{2}\nonumber\label{h0r}\\
&  +i\alpha_{h}e^{3i\varphi}\left(  F_{0}+F_{1}\partial_{\varphi}%
+F_{2}\partial_{\varphi}^{2}+F_{3}\partial_{\varphi}^{3}\right)  \sigma
_{-}\nonumber\\
&  -i\alpha_{h}e^{-3i\varphi}\left(  F_{0}-F_{1}\partial_{\varphi}%
+F_{2}\partial_{\varphi}^{2}-F_{3}\partial_{\varphi}^{3}\right)  \sigma_{+},
\end{align}
where $F_{0}=i(R_{0}-3R_{1}+3R_{2})$, $F_{1}=-3R_{1}+9R_{2}-8R_{3}$,
$F_{2}=i(-3R_{2}+6R_{3})$, and $F_{3}=R_{3}$. $R_{i}$s
depend on the radial confinement and are defined in Section \ref{holesDress}.

The current operator in spin space is:
\begin{align}
\frac{\hat{I}_{l}^{\varphi}}{e}= &  2E_{a}^{h}\left(  l_{-}^{\prime}%
 {\sigma}_{1}+l_{+}^{\prime} {\sigma}_{2}\right)  \nonumber\\
&  -\frac{\alpha_{h}}{\hbar}\left(  F_{1}+2iF_{2}l_{-}^{\prime}+3F_{3}%
l_{-}^{\prime}{}^{2}\right)  \sigma_{-}\nonumber\\
&  -\frac{\alpha_{h}}{\hbar}\left(  F_{1}-2iF_{2}l_{+}^{\prime}+3F_{3}%
l_{+}^{\prime}{}^{2}\right)  \sigma_{+},
\end{align}
where $l_{+}^{\prime}=l+1$, and $l_{-}^{\prime}=l-2$. Thus, the same procedure as before, leads to
the Hamiltonian carrying a ground state current. In spin
space:
\begin{align}
{H}^{cR}= &  {H}_{0}^{cR}\left(  l-n_{\lambda}\right)  |_{\bar{F}%
_{1}\rightarrow F_{1},\bar{F}_{0}\rightarrow F_{0}}\label{Hjcd2}\\
&  -2\alpha_{h}e^{i\varphi}F_{3}n_{\lambda}^{3}\sigma_{-}+2\alpha_{h}%
F_{3}n_{\lambda}^{3}e^{-i\varphi}\sigma_{+},\nonumber
\end{align}
where $\bar{F_{1}}=F_{1}+3F_{3}n_{\lambda}^{2}$, and $\bar{F_{0}}=F_{0}%
-F_{2}n_{\lambda}^{2}$, and we disregard a constant shift of $-\hbar
^{2}n_{\lambda}^{2}/(2m_{hh}a^{2})$ in the Hamiltonian.
Eigenstates are now:
\begin{align}
&  \psi_{(l,n_{\lambda}),+}^{cR}=e^{il\varphi}\left(
\begin{array}
[c]{c}%
\cos\frac{\theta^{cR}(l,n_{\lambda})}{2}e^{-(3i/2)(\varphi)}\\
\sin\frac{\theta^{cR}(l,n_{\lambda})}{2}e^{(3i/2)(\varphi)}%
\end{array}
\right)  ;\\
&  \psi_{(l,n_{\lambda}),-}^{cR}=e^{il\varphi}\left(
\begin{array}
[c]{c}%
-\sin\frac{\theta^{cR}(l,n_{\lambda})}{2}e^{-(3i/2)(\varphi)}\\
\cos\frac{\theta^{cR}(l,n_{\lambda})}{2}e^{(3i/2)(\varphi)}%
\end{array}
\right)  ,
\end{align}
where the texture angle $\theta^{cR}(l,n_{\lambda})$ is
\begin{equation}
\theta^{cR}(l,n_{\lambda})=\tan^{-1}\left[  \tilde{\alpha}_{h}\left(  \frac
{2}{3}R_{0}+\left(  \frac{13}{12}-\frac{1}{3}l^{2}+n_{\lambda}^{2}-\frac{2}%
{3}\frac{n_{\lambda}^{3}}{l}\right)  R_{3}\right)  \right]  ,\label{bn2}%
\end{equation}
with $\tilde{\alpha}_{h}=2m_{hh}\alpha_{h}/(\hbar^{2}R_{3}^{2/3})$ and the energies
\[
E_{k,\sigma}=E_{a}\left(  \left(  l-n_{\lambda}\right)  ^{2}+\frac{1}%
{4}+\sigma\frac{\left(  l-n_{\lambda}\right)  }{\cos\left[  \theta^{cR}\left(
\left(  l-n_{\lambda}\right)  ,n_{\lambda}\right)  \right]  }\right)  .
\]
For small $\theta^{cR}$ we find, as above,
\begin{equation}
n_{\lambda}\approx-\frac{\hbar}{4E_{a}^{h}}\frac{\pi}{e}\frac{I}{n_{r}}.
\end{equation}
and a spin polarization in the $z$-direction:
\begin{equation}
\left\langle \mathbf{\sigma}_{z}\right\rangle _{I_{\varphi}}^{cR}\approx
-\frac{\epsilon_{F}}{E_{a}^{h}}\frac{\pi\hbar\alpha_{h}^{2}}{9\left(
E_{a}^{h}a^{2}\right)  ^{3}}\frac{I}{e}.\label{SZFr}%
\end{equation}
very similar to the Dresselhaus limit, but with a prefactor $1/9$. Therefore
above discussions for the small SOI limit hold for the cubic Rasba Hamiltonian
as well. In the lower part of Fig. \ref{numeric}, we plotted numerical results
for larger values of $\alpha_{h}$. The values of the SO coupling used in this figure can
be experimentally achieved (e.g.\ Ref.\ \onlinecite{Pfeuffer}),
by external gate voltage.

\section{Conclusions\label{con}}

IFE allows in principle ultrafast and non-dissipative actuation and
eventual switching of magnetization. We investigated the impact of the SOI on
this non-absorbing `Opto-Spin' phenomena. We provided a proof of principle for a
mechanism that is based on the current-induced generation of a spin
polarization that would generate torques in a magnetic sample. The current
bias can be generated by the Lagrange multiplier method
inspired by current-density functional theory. For electrons moving in
quantum rings in the presence of the Rashba and the Dresselhaus SOIs, the effect vanishes.
It becomes non-zero only when the Kramers' degeneracy is broken by an exchange
potential or applied magnetic field, but the effects are still
small.\cite{Joibari} On the other hand, holes in a ring with cubic
Dresselhaus and Rashba SOI display spin polarization under current bias.
This polarization is a competition between two effects. On one hand, with
increasing SOI the band splitting increases, which amplifies the magnitude of
the polarization. Simultaneously, however, the $z$-component of the spin of
electrons with energies near to the Fermi level decreases, and therefore the net
polarization decreases. These two might enhance the effect rather than
cancel each other when the spin texture would push the spin out of the plane. This
can be achieved in a
ring with an asymmetric potential in the radial direction, such as a thin
slice of GaAs%
$\vert$%
p-doped GaAlAs core/shell nanowire. The second Rashba SOI would pull the spin
toward the $z$-direction and lead to monotonic increase of current-induced
spin polarization with SOI.

Induced polarization in the z-direction, calculated in this Article, could be
either parallel or anti-parallel to the $z$-axis depending on the direction of
the current. This is consistent with the IFE in which the effective
magnetic field changes sign with the helicity of light. Here, we focused
on the spin polarization induced by current in a material which is
nonmagnetic. This spin polarization can be measured directly by pump and Kerr
rotation probe measurements. GaMnAs in the ferromagnetic state is a hole
conductor. Here the current-induced spin polarization would induce torques on
the magnetic order parameter, eventually causing magnetization switching. The
spin-dependent dynamic Stark effect also induces torques by circularly
polarized light.\cite{Qaium} The two processes are independent and should be
added. They can be distinguished by tuning the light frequency close to the
energy gap, where the dynamic Stark effect is resonantly enhanced.

The currents generated by non-resonant light are persistent, analogous to the
diamagnetic currents in conducting rings induced by dc magnetic
fields.\cite{Buttiker} While this issue has not been central to our study, our
results imply that the spin-orbit interaction can induce large paramagnetic
corrections to the diamagnetic response. Cantilever-based torsional
magnetometers with integrated mesoscopic rings allow very sensitive
measurements of magnetic susceptibilities.\cite{Harris} We suggest that
quantum ring arrays made from 2DHGs would be interesting subjects for such experiments.

\begin{acknowledgements}
This work was supported by FOM (Stichting voor Fundamenteel Onderzoek der
Materie){,}\ by the ICC-IMR, by DFG Priority Programme 1538 \textquotedblleft%
{Spin-Caloric Transport}\textquotedblright\ (GO 944/4), by KAKENHI (Grants-in-Aid for Scientific Research)
Nos. 25247056, and 25220910, and by the National Science Foundation under Grant No. NSF PHY11-25915.
\end{acknowledgements}

\appendix

\section{Light induced currents\label{appe}}

Here we show how to use the collisionless plasma model by Hertel\cite{Hertel}
to obtain the light-induced current in a quantum ring. This model can be
used for the present system in the high-frequency limit, in which the path an
electron traverses under a half-cycle of the oscillating light electric field
is much smaller than the characteristic length scales such as the finite
radial thickness or the spin-orbit precession length.

Hertel finds a circular current as a result of the circularly polarized light
in the form of
\begin{equation}
\mathbf{j}_{\varphi}=-\frac{i}{4e\langle n\rangle\omega}\boldsymbol{\nabla
}\times\left[  \sigma^{\ast}{\mathbf{E}}^{\ast}\times\sigma{\mathbf{E}%
}\right] \ ,
\end{equation}
where ${\mathbf{E}}$ is the electric field of the light, and
\begin{equation}
\sigma=\frac{i\langle n\rangle e^{2}}{m\omega},
\end{equation}
is the conductivity of a collisionless plasma in a high frequency regime,
$\langle n\rangle$ is the volume density of the electrons and $\omega$ is the
light frequency. For circularly polarized light with helicity $\Lambda=\pm$
\[
{\mathbf{E}}\times{\mathbf{E}}^{\ast}=\Lambda i\left\vert E\right\vert
^{2}\cdot\mathbf{e}_{z} \ . %
\]
Thus,
\begin{align}
\mathbf{j}_{\varphi}(r)  &  =\Lambda\frac{\langle n\rangle e^{3}}{4m^{2}%
\omega^{3}}\boldsymbol{\nabla}\times\left(  \left\vert E\left(  r\right)
\right\vert ^{2}\mathbf{e}_{z}\right) \\
&  =-\Lambda\frac{\langle n\rangle e^{3}}{4m^{2}\omega^{3}}\left(
\frac{\partial\left\vert E\left(  r\right)  \right\vert ^{2}}{\partial
r}\right)  \mathbf{e}_{\varphi} \ . %
\end{align}
Since this result does not depend on the $z$-coordinate, it holds for 2DEG
and normally incident light.

In a ring we can project the current to one dimension by writing the current
density
\begin{equation}
\mathbf{j}_{\varphi}=\left\vert R_{0}(r)\right\vert ^{2}\left\vert
Z_{0}(z)\right\vert ^{2}J_{ext}^{1D}\mathbf{e}_{\varphi}, \label{jphi}%
\end{equation}
where%
\begin{align}
J_{ext}^{1D}  &  =\left\langle R_{0}(r)Z_{0}(z)\right\vert {j}_{ext}\left(
{r}\right)  \left\vert Z_{0}(z)R_{0}(r)\right\rangle \\
&  =-\Lambda\frac{Ne^{3}}{8\pi am^{2}\omega^{3}}\int dr\left\vert
R_{0}(r)\right\vert ^{2}\frac{\partial\left\vert E\left(  r\right)
\right\vert ^{2}}{\partial r} \ ,
\end{align}
where we used $\int dr\int dz\langle n\rangle=N/(2\pi a)$, the linear density
of a ring with $N$ electrons. We consider a laser spot with Gaussian spatial
distribution:
\begin{equation}
\mathbf{E}(r)=(\mathbf{e}_{x}+\Lambda i\mathbf{e}_{y})E_{0}\exp\left(
-\frac{\gamma r^{2}}{2}\right)  ,
\end{equation}%
\begin{equation}
\left\vert E\right\vert ^{2}=E_{0}^{2}\exp\left(  -\gamma r^{2}\right)  ,
\end{equation}
where $E_{0}$ is the maximum value of the electric field in the spot center.
The current density in the ring with radius $a$ then becomes
\begin{equation}
J_{ext}^{1D}=\Lambda\gamma E_{0}^{2}\exp\left(  -a^{2}\gamma\right)
\frac{Ne^{3}}{4\pi m^{2}\omega^{3}}.%
\end{equation}
The above current has dimension of Ampere. The result is also valid for the
holes (with the modified mass and the opposite current direction). The light intensity
reads in terms of the electric field
\begin{equation}
\text{Intensity}=\frac{cn^{\prime}\epsilon}{2}\left\vert E\right\vert ^{2},
\end{equation}
where $c$ is the velocity of light in vacuum, $\epsilon$ is the dielectric
constant, and $n^{\prime}$ the index of refraction. We estimate the
current by assuming $\epsilon\approx10\epsilon_{0},$ $n^{\prime}\approx
3$\textit{. }At a typical laser intensity of $10^{13}\,\mathrm{Wm}^{-2\text{
}}$ and frequency used in all-optical switching\cite{vah} $|E_{0}|^{2}%
\approx3\times10^{15}\,\mathrm{V}^{2}\mathrm{m}^{-2}$ and wave
length/frequency $\lambda^{\prime}=12\pi\,\mathrm{\mu m\,}$\textrm{/}%
$\mathrm{\,}\omega=2\times10^{14}\mathrm{s}^{-1}$ we find for the current in a
2DHG ring: \textit{ }
\begin{align}
\left\vert I\right\vert =16\,\mathrm{nA}  &  \frac{E_{0}^{2}}{3\times
10^{15}\mathrm{V}^{2}\mathrm{m}^{-2}}\frac{\gamma\exp\left(  -a^{2}%
\gamma\right)  }{10^{12}\exp(-1)\mathrm{m}^{-2}}\frac{N}{1000}\\
&  \times\left(  \frac{0.45m_{0}}{m}\right)  ^{2}\left(  \frac{2\times
10^{14}\mathrm{s}^{-1}}{\omega}\right)  ^{3} \ . %
\end{align}

\end{document}